\documentstyle[preprint,prd,aps,epsf,psfrag,graphicx,latexsym]{revtex}
\begin{document}
\newcommand{\tkl}{{{\small\sf T}{$\chi$}{\small\sf L}} }
\newcommand{\tkll}{{{\sf T}{$\chi$}{\sf L}}}
\newcommand{\sesaml}{{\sf SESAM}}
\newcommand{\sesam}{{\small\sf SESAM }}
\newcommand{\sesams}{{\small\sf SESAM}'s }
\newcommand{\sesamc}{{\small\sf SESAM} collaboration}
\newcommand{\largepspicture}[1]{\centerline{\setlength\epsfxsize{10cm}\epsfbox{#1}}}
\newcommand{\vlargepspicture}[1]{\centerline{\setlength\epsfxsize{15cm}\epsfbox{#1}}}
\newcommand{\smallpspicture}[1]{\centerline{\setlength\epsfxsize{8.0cm}\epsfbox{#1}}}
\newcommand{\tinypspicture}[1]{\centerline{\setlength\epsfxsize{7.4cm}\epsfbox{#1}}}
\newcommand{\ewhxy}[3]{\setlength{\epsfxsize}{#2}
            \setlength{\epsfysize}{#3}\epsfbox[0 20 660 580]{#1}}
\newcommand{\ewxy}[2]{\setlength{\epsfxsize}{#2}\epsfbox[10 30 640  590]{#1}}
\newcommand{\ewxynarrow}[2]{\setlength{\epsfxsize}{#2}\epsfbox[10 30 560 590]{#1}}
\newcommand{\ewxyvnarrow}[2]{\setlength{\epsfxsize}{#2}\epsfbox[10 30 520 590]{#1}}
\newcommand{\ewxywide}[2]{\setlength{\epsfxsize}{#2}\epsfbox[0 20 380 590]{#1}}
\newcommand{\err}[2]{\raisebox{0.08em}{\scriptsize{$\hspace{-0.8em}\begin{array}{@{}l@{}}
                     \plus\makebox[0.55em][r]{#1}\\[-0.15em]
                     \minus\makebox[0.55em][r]{#2}
                     \end{array}$}}}
\newcommand{\plus}{\makebox[15pt][c]{$+$}}
\newcommand{\minus}{\makebox[15pt][c]{$-$}}
\newcommand{\mpr}{\frac{m_{\pi}}{m_{\rho}}}
\newcommand{\beq}{\begin{equation}}
\newcommand{\eeq}{\end{equation}}
\newcommand{\plb}{Phys.~Lett.~B}
%
\renewcommand{\baselinestretch}{1.25}
%
%

\preprint{HLRZ 1997\_35, WUB 97-24} 

\title
{Improved $\Upsilon$ Spectrum with Dynamical Wilson Fermions.}

\author{N.~Eicker, Th.~Lippert, K.~Schilling, A.~Spitz\thanks{email:
  spitz@hlrz.kfa-juelich.de}}

\address{HLRZ, Forschungszentrum J\"ulich, 52425 J\"ulich and\\
DESY, 22603 Hamburg, Germany}

\author{J.~Fingberg, S.~G\"usken, H.~Hoeber\thanks{email:
    hoeber@theorie.physik.uni-wuppertal.de}, J.~Viehoff}
\address{Fachbereich Physik, Bergische Universit\"at, Gesamthochschule
Wuppertal\\Gau\ss{}stra\ss{}e 20, 42097 Wuppertal, Germany}

\author{\sesamc}

\date{\today}

\maketitle

\tighten
\begin{abstract}
We present results for the $b \bar b$ spectrum obtained using an 
${\cal O}(M_bv^6)$-correct non-relativis\-tic lattice QCD action, where
$M_b$ denotes the bare b-quark mass and $v^2$ is the mean squared
quark velocity. Propagators are evaluated on \sesams three sets of
dynamical gauge configurations 
generated with two flavours of Wilson fermions at $\beta = 5.6$. These
results, the first of their kind obtained with dynamical Wilson fermions, are
compared to a quenched analysis at equivalent lattice spacing, $\beta
= 6.0$. Using our three sea-quark values we perform the ``chiral''
extrapolation to $m_{\rm eff} = m_s/3$, where $m_s$ denotes the strange quark
mass. The light quark mass dependence is found to be small in relation
to the statistical errors. Comparing the full QCD result to our quenched
simulation we find better agreement of our dynamical data with
experimental results in the spin-independent sector but observe no
unquenching effects in hyperfine-splittings. To pin down the
systematic errors we have also compared quenched results in
different ``tadpole'' schemes as well as using a lower order action. We find
that spin-splittings with an ${\cal O}(M_bv^4)$ action are ${\cal
  O}$(10 \%) higher compared to ${\cal O}(M_bv^6)$ results. Relative
to the results obtained with the plaquette method the Landau gauge
mean link tadpole scheme raises the spin splittings by about the same
margin so that our two improvements are opposite in effect.
\end{abstract}
\newpage
\clearpage
\section{Introduction}
Not just since the formulation of Heavy Quark Effective Theory a few years
ago \cite{hqet}, systems containing one or two heavy quarks have been the focus of
much attention, both theoretically and experimentally. Today, B
mesons are at the center of experimental efforts to determine
quantities related directly to the CKM matrix, always in the hope that
new physics might emerge. But, just as for systems with light quarks,
the properties of hadrons containing heavy quarks are determined by
non-perturbative physics and here the lattice can provide the missing
link. 
\par
Non-relativistic QCD (NRQCD) \cite{bod} is an 
effective theory used to describe heavy quarks, $M_q \gg \Lambda_{\sf
  QCD}$. The NRQCD Lagrangian is written as a series of operators
expanded in powers of the mean squared heavy quark velocity $v^2$, ${\cal
  L}_{\sf NRQCD} = \sum_{i,n} c_i(g^2) O_i(M_q v^{2n})$ where the
coefficients $c_i$ are obtained by perturbative matching with QCD. The
lattice version of NRQCD \cite{lepage,lepthack} allows one to simulate
heavy quarks with lattice spacing errors of ${\cal O}(\vec p
a)$, where $p$ is a typical momentum $\approx$ $\Lambda_{\sf QCD}$, 
rather than the usual ${\cal O}(M_q a)$ for Wilson-type heavy
quarks. This makes lattice NRQCD a promising technique to
simulate systems containing a $b$-quark.
\par
Lattice NRQCD has been widely used in the past few years to calculate a
variety of phenomenologically interesting quantities (see
\cite{christine_tsuk} for a recent summary). The first step in the
NRQCD program, naturally, is the calculation of the spectrum of the $b
\bar b$ system. Such calculations have been performed using the 
quenched approximation by \cite{nrcam,davies,manke}. So far, there has
only been one analysis of the $\Upsilon$ with dynamical quarks: the
authors of \cite{christine_tsuk,staggered,alpha} applied an ${\cal
  O}(M_bv^4)$ correct NRQCD action to gauge fields incorporating the
effects of two dynamical staggered fermions\footnote{The high
  statistics measurements in the spin-independent sector in
  \cite{staggered,alpha} were obtained using an ${\cal O}(M_bv^2)$
  correct action.}. It was found that the experimentally known 
{\it spin-independent} spectrum could not be reproduced using the
quenched approximation whereas the data with 2 dynamical fermions
agrees much better. These measurements were pushed to high statistics
(4\% statistical errors) in \cite{staggered,alpha} and used to obtain
precision measurements of the strong coupling constant $\alpha_s$. 
Quenching, the approximation in which the effects of dynamical
fermions are disregarded, was concluded to be the largest source of
uncertainty in this calculation. The dependence of the splittings 
on the dynamical quark {\it mass} was estimated, using just two mass
values, to be of the same order as the statistical error. No
simulation exists, so far, applying NRQCD to dynamical {\it Wilson}
fermions. 
\par
In the {\it spin-dependent} sector the emerging picture is much less
clear. The P-fine structure in the unquenched theory \cite{christine_tsuk,alpha},
obtained with the ${\cal O}(M_bv^4)$-correct action, was found to be
in very good agreement with experiment whereas the quenched results,
using the same action, predict much smaller splittings. However,
recent results of \cite{manke,scaling,trottier} (all of which use the quenched
approximation) have exposed the sensitivity of the spin-dependent
splittings to the details of the action: several improvements, such as
the inclusion of higher order spin-dependent terms in the NRQCD
expansion, the addition of lattice-artefacts correcting terms of
${\cal O}(a^2)$ \cite{manke,trottier} as well as an improved phenomenological estimate of
the coefficients $c_i$ \cite{trottier,lep_tsuk} were found to have
sizeable effects of ${\cal O}(10-20 \%)$ for these 
splittings. Thus, there has recently been strong support in favour of
using the ${\cal O}(M_bv^6)$-correct NRQCD action, where, at highest
order, all {\it spin-dependent} corrections are added, as well as for
the use of the mean link Landau gauge tadpole scheme to estimate the perturbative
coefficents $c_i$. The effects of these improvements for dynamical
configurations have not been studied.
\par
In the simulation presented here we attempt to address many of the
open issues pin-pointed above. Using \sesams large sample of
dynamical Wilson-fermion gauge configurations we study both the
spin-independent as well as the spin-dependent spectrum of the
$\Upsilon$. Our strategy, in searching for sea-quark effects, will be
to compare our final dynamical results to that of a quenched
simulation at equivalent lattice spacing. Thus, we hope to consolidate that
unquenching brings the spin-independent splittings into good agreement
with experiment. Using our three sea-quark masses we can also study the
dependence of mass-splittings on the light sea-quark mass. Following
the recent suggestions of \cite{manke,trottier} we have implemented
the NRQCD action including spin-dependent corrections of ${\cal O}(M_bv^6)$
and we remove tadpoles using the mean link calculated in Landau
gauge. With these ingredients we hope to clarify the effect of
unquenching in the spin-dependent splittings. In addition, using
the quenched configurations we investigate: (i) the
effect of changing tadpole prescriptions; (ii) the effect of using an
${\cal  O}(M_bv^4)$ correct action compared to the ${\cal O}(M_bv^6)$
corrected one. 
\section{Simulation Details}
In section \ref{secact} we present the discretized version of
our ${\cal O}(M_bv^6)$ correct NRQCD action. Section \ref{sectad}
addresses issues concerning the determination of the perturbative
coefficients $c_i(g^2)$. 
\subsection{Action\label{secact}}
The non-relativistic (Euclidean) lattice Hamiltonian to ${\cal
  O}(M_bv^6)$ consists of \cite{lepage}: the kinetic energy operator, 
\begin{equation}
  H_0 = -\frac{\Delta^{(2)}}{2M_b} \; ,
\end{equation}
which is of order $M_bv^2$; relativistic corrections of order $M_bv^4$,
\begin{eqnarray}
  \delta H^{(4)} & = & -c_1\frac{\left(\Delta^{(2)}\right)^2}{8M_b^3} +
  c_2\frac{ig}{8M_b^2}\left(\boldmath\Delta\cdot E -
  E\cdot\boldmath\Delta\right) \nonumber \\
  && - c_3 \frac{g}{8M_b^2}\sigma\cdot\left( \tilde\Delta\times\tilde {\bf E} -
  \tilde{\bf E}\times\tilde\Delta\right) - c_4\frac{g}{2M_b}\sigma\cdot
  \tilde{\bf B} \nonumber \\
  && + c_5 \frac{a^2\Delta^{(4)}}{24M_b} - c_6
  \frac{a\left(\Delta^{(2)}\right)^2}{16nM_b^2} \; ,
\end{eqnarray}
and spin-sensitive corrections of order $M_bv^6$, 
\begin{eqnarray}
  \delta H^{(6)} & = & - c_7 \frac{g}{8M_b^3}\lbrace
  \Delta^{(2)},\sigma\cdot {\bf B}\rbrace \nonumber \\
  && - c_8 \frac{3g}{64M_b^4} \lbrace
  \Delta^{(2)},\sigma\cdot\left(\Delta\times{\bf E} - {\bf
  E}\times\Delta\right)\rbrace \nonumber \\
  && - c_9 \frac{ig^2}{8M_b^3}\sigma\cdot{\bf E}\times {\bf E} \; .
\end{eqnarray}
We do not include spin-independent corrections of ${\cal
  O}(M_bv^6)$. The $E$ and $B$ fields are definded in terms of the
  field strength tensor
\begin{eqnarray}
F_{\mu\nu}(x) &=& - \frac{1}{4} \sum_{\Box}\left( \frac{U_{\mu
      \nu}(x) - U_{\mu\nu}^{\dagger}(x)}{2 i} - 
\frac{1}{3} {\sf Tr}({\sf Im} U_{\mu\nu}(x)) \right) \; , \nonumber  \\
  E^{i} & = & F^{i0} \nonumber \; , \\
  B^{i} & = & -\frac{1}{2}\epsilon_{ijk}F^{jk} \; , \label{fmunu}
\end{eqnarray}
where, $U_{\mu\nu}$ is the standard plaquette and the sum is over all
anticlockwise plaquettes in the $\mu$-$\nu$-plane, $1 \leq \mu \leq
4$, $1 \leq \nu < \mu$.
\par
The removal of ${\cal O}\left(a^2 M_b v^4\right)$ discretization
errors is achieved by: (i) using
an improved version of the lattice field strength tensor \cite{lepage}
in the leading spin-dependent interactions ($c_3$ and $c_4$): 
\begin{eqnarray}
\tilde{F}_{\mu \nu} &=& \frac{5}{3} F_{\mu \nu} - \frac{1}{6}
\left(U_\mu(x) F_{\mu \nu} (x + \hat{\mu}) U_{\mu}^{\dagger}(x) \right.
  \nonumber \\
&& \left. + U_{\mu}^{\dagger}(x - \hat{\mu})F_{\mu \nu}(x - \hat{\mu}) U_\mu(x -
  \hat{\mu}) - (\mu \leftrightarrow \nu) \right)  \; ;
\end{eqnarray}
(ii) adding the last two terms in $\delta H^{(4)}$ to (classically)
  correct for finite lattice spacing errors in the spatial and 
  temporal derivatives and (iii) by using the improved derivative operator
\begin{eqnarray}
\tilde{\Delta}_i &=& \Delta_i - \frac{a^2}{6} \Delta_i^{(+)} \Delta_i
\Delta_i^{(-)} \\
a \Delta_i^{(+)}G(x) &=& U_i(x) G(x + a \hat i) - G(x) \nonumber \\
a \Delta_i^{(-)}G(x) &=& G(x) - U_i^{\dagger}(x - a \hat i)G(x - a
\hat i) \nonumber \\
a \Delta_i G(x) &=& \frac{1}{2} \left(U_i(x) G(x + a \hat i) -
  U_i^{\dagger}(x - a \hat i)G(x - a \hat i) \right) \nonumber
\end{eqnarray}
in the leading spin-dependent interactions ($c_3$).
\par
For completeness we also give our definitions of $\Delta^{(2)}$ and 
$\Delta^{(4)}$:
\begin{eqnarray}
\Delta^{(2)} &=& \sum_i \Delta_i^{(2)} \nonumber \\
a^2 \Delta_i^{(2)} G(x) &=& U_i(x) G(x + a \hat i) + U_i^{\dagger}(x -
a \hat i) G(x - a \hat i) - 2 G(x) \nonumber \\
\Delta^{(4)} &=& \sum_i \left( \Delta_i^{(2)} \right)^2 \, .
\end{eqnarray}
\par
Following \cite{davies} the quark Greens functions are calculated from the
evolution equation 
\begin{eqnarray}
  G(t+1) & = & \left( 1 - \frac{aH_0}{2n}\right)^n U_4^{\dag}\left( 1 -
  \frac{aH_0}{2n}\right)^n\left( 1 - a\delta H \right) G(t), \\
  G(0) & = & \delta_{{\bf x},0} \; , \nonumber
\end{eqnarray}
where $G(t)$ is a two-spinor component object.
\par
The parameter $n$ serves to stabilise the evolution for small
bare quark masses and is set to $n=2$ \cite{lepage}. 
\subsection{Tadpoles\label{sectad}}
An important feature of our simulation is the use of
``tadpole''-improvement: gauge links $U_\mu$ are replaced by $U_\mu /
u_0$ where the mean link $u_0$ accounts for the effects of tadpole
diagrams \cite{parisi,lepmac}. The coefficients $c_i$ are then set to
their tree-level values of 1. 
\par
The choice of $u_0$ is, of course, not unique; here we choose $u_0$ to
be the mean value of the link variable in the Landau gauge:
\begin{equation}
u_{0}^{(L)} = \langle \frac{1}{3} {\sf Tr}
U_{\mu} \rangle \, , \hspace{2cm} \partial_\mu A_\mu = 0 \, .
\end{equation}
With this choice of gauge $\langle  {\sf Tr}
U_{\mu} \rangle $ is maximised so that $u_0$ is as close to 1 as
possible; it was therefore argued in \cite{lep_tsuk} that remaining
tadpole contributions cannot be due to lattice artefacts. 
A recent NRQCD lattice calculation of the $c
\bar c$ spectrum over a variety of 
lattice spacings suggests that this choice improves the scaling
behaviour of spectroscopic quantities \cite{trottier}, notably the
hyperfine splittings. This view is also endorsed in
\cite{christine_tsuk,scaling} where the authors have studied the scaling
behaviour in the quenched $b \bar b$ spectrum using three values of the
coupling ($\beta = 5.7,6.0,6.2$), an ${\cal O}(M_bv^4)$-correct action
and the plaquette tadpole value (defined below). They find the
spin-independent spectrum to display insignificant scaling violations
whereas the hyperfine splittings do not scale that well. This is
attributed partly to ${\cal O}(a^2)$ errors in the $B$
field\footnote{which was obtained fom the unimproved field strength
  tensor, eq.\ref{fmunu}.} arising
in the term $\sigma \cdot B$ but partly also to the choice of
plaquette tadpole
improvement not capturing the tadpole effects sufficiently
well. Additionally, two more arguments in favour of
$u_{0}^{(L)}$ are given in \cite{lep_tsuk}: firstly, the static
potential shows less violation of rotational invariance using a
$u_{0}^{(L)}$-tadpole improved gluonic action and, secondly, the
non-perturbative determination of the clover coefficient \cite{luscher} is in good
agreement with the $u_{0}^{(L)}$-tadpole improved perturbative result. 
Since, for the $b \bar b$, all
simulations performed to date, notably those with dynamical staggered
quarks, have been using $u_{0}^{(P)} = \langle  \frac{1}{3} {\sf Tr}
U_{\mu \nu} \rangle^{\frac{1}{4}}$, our measurement will help to
explore the size of the systematic error associated with the choice of
$u_0$. In the quenched sector we have also performed an analysis of
the S-splittings (where signals are cleanest) using the plaquette
prescription and, furthermore, we compare the ${\cal O}(M_b v^6)$-results
to a simulation where the NRQCD action is correct to ${\cal O}(M_b v^4)$. 
In table \ref{tadpole} we compare our values of $u_{0}^{(L)}$ to the commonly used
$u_{0}^{(P)}$. Note that $u_0$ weights the field strength tensor with
four powers so that, naively, using $u_{0}^{(P)}$ instead of
$u_{0}^{(L)}$ may change the hyperfine splittings by as much as 8\%. 
\subsection{Correlators and Smearing}
Meson correlation functions at zero momentum are built from quark propagators combined with
suitable interpolating operators:
\begin{equation}
  G^{\rm meson}_{\rm sc,sk}\left( t, {\bf p} = 0\right) = \sum_{\bf x,y} {\rm Tr} \left[
  G^{\dag}\left({\bf x},t\right)\Gamma_{(sk)}^{\dag}\left({\bf y -
  x}\right)\tilde G\left({\bf y},t \right)\right] \; , 
\end{equation}
where the source smeared propagator $\tilde G$  is obtained by evolving the
extended source:
\begin{equation}
  \tilde G\left({\bf y}\right) = \sum_{\bf x} G\left({\bf y -
  x},t\right) \Gamma_{(sc)}\left({\bf x}\right) \; .
\end{equation}
The interpolating operator 
$\Gamma^{(sc/sk)}({\bf x}) \equiv \Omega
\Phi^{(sc/sk)}\left({\bf x}\right)$ contains a spin matrix $\Omega$ and a
spatial smearing function $\Phi$; the superscript $sc/sk$ denotes
source and sink smearing respectively. A proper choice of $\Phi$ is
crucial to obtain clear signals for excited states. Here, we benefit
from a recent lattice determination of the $b \bar b$ potential
including relativistic corrections \cite{smearing}. These authors
start from the ${\cal O}(v^4)$ non-relativistic Schr\"odinger-Pauli Hamiltonian
\cite{pauli}, which, in the center of mass frame of the quark and
antiquark, has the form
\begin{eqnarray}
H &=& \sum_{i=1,2} \left( m_i + \frac{p ^2}{2m_i} - \frac{p^4}{8 m_i^3} +
\frac{1}{8 m_i^2} \nabla^2(V_0(r) + V_a(r)) \right) \nonumber \\
&& + V_0(r) + V_{\rm
sd}(r,\vec L, \vec S_1, \vec S_2) + V_{\rm vd}(r, \vec p)\, .
\end{eqnarray}
It contains the central potential of Cornell type:
\begin{equation}
V_0(r) = V + k r - \frac{e}{r} \, ,
\end{equation}
spin- and velocity-dependent (sd, vd) potentials (see \cite{smearing})
and the Darwin-term $V_a$. Following \cite{smearing} we numerically
integrate the radial Schr\"odinger equation
\begin{equation}
g''_{E,L}(r) = F(r,E,L,\alpha,m, k) \, g_{E,L}(r)
\end{equation}
for definite radial quantum number and angular momentum, where
\begin{equation}
F(r,E,L,\alpha,m, k) = \frac{L(L+1)}{r^2} - m\left( E - k r +
  \frac{\alpha}{r} \right) \, .
\end{equation}
The term $-\frac{\alpha}{r}$ contains the Coulomb-term of the central
potential, the $\frac{1}{r}$ contribution of the Darwin term as well
as the leading $\frac{1}{r}$ contribution from $V_{\rm vd}$. We leave
$m$ and $k$ as free parameters but fix the contributions of the terms
not coming from the central potential to the lattice values found in
\cite{smearing}. The Coulomb coefficient $e$ is that of the quenched
simulation in \cite{smearing} {\it augmented} by 20 \% to take into
account the different running at short distances observed in
\cite{sesampot}. A direct comparison with hydrogen wave functions
clearly demonstrates the superiority of our choice.
\par
The continuum wave functions are converted to the lattice using the
scales obtained from \sesams $\rho$-masses for each $\kappa_{\rm
sea}$. The spin matrices $\Omega$ are taken from
\cite{davies} and listed in table \ref{operators}.
Note that the finite difference operators included in table
\ref{operators} for the P-waves are only applied in the case of
delta-function smearing. 
\par
For the $\Upsilon$ and 
$\eta_b$ we calculate a $4\times 4$ matrix of correlators with four
different smearings at source and sink, $sc/sk
= local,1,2,3 $ corresponding to a point source (local), the ground
state (1), the first (2) and second (3) excited states
respectively. For the $L=1$ states we restrict ourselves to the ground
state and the first excitation as signals deteriorate. Correlators
with momenta upto $|{\bf p}| = 2$ are also calculated. Gauge
configurations are fixed to Coulomb gauge.
\subsection{Lattice Parameters}
The lattice parameters we have used are displayed in table
\ref{simulation}. 
%
%
\sesam has recently completed the generation of
gauge configurations with two (degenerate) dynamical fermions at three
values of the sea-quark mass. For each mass the run consists of ${\cal
  O}(5000)$ trajectories\footnote{In a standard Hybrid Monte Carlo
  algorithm we typically set the time-step size $dt = 0.01$ and
  perform 100 molecular dynamics updates to generate one
  trajectory.} and we analyse every 25th trajectory giving a total of
200 configurations at each sea-quark. This is by far
the largest set of dynamical Wilson-fermion gauge configurations to date,
more than a factor two larger than the sample analysed with respect to
NRQCD in
\cite{staggered}\footnote{The authors of \cite{staggered} analyse 400
  configurations taken from a total trajectory length of 2000.}. Details concerning
the generation of our dynamical configurations and issues surrounding
autocorrelations are discussed in ref.~\cite{sesamauto}. Here, we
mention only the values of $\pi$ to $\rho$ mass ratio and the lattice
spacing obtained from the $\rho$ - which we have measured in a
standard light spectrum calculation on the lattice
\cite{sesamlight,sesamspec} - for our three dynamical sea quarks
$\kappa_{\rm sea} = \{0.156, 0.157, 0.1575 \}$: $\frac{m_\pi}{m_\rho} = \{ 0.839(4), 0.755(7), 0.69(1) \}$ and $a^{-1}_\rho = 2.33(6) {\sf GeV}$.
\par
Taking advantage of the smallness of the
bottomonium system, we exploit configurations more than once
by starting the propagator evolution both at different spatial source
points and on different timeslices. A binning procedure confirms that
our 4 measurements per configuration are indeed independent.
\par
Errors are obtained from correlated fits using the bootstrap
procedure. They correspond to 68 \% confidence limits of the
distribution obtained from 300 bootstrap samples.
\par
Throughout, we fix the bare heavy quark mass to a value $aM_b =
1.7$. In a quenched simulation by the NRQCD collaboration at $\beta =
6.0$,  a bare $b$ quark mass of 1.71 was found to yield the correct
physical value of the $\Upsilon$ \cite{davies_mb}. 
%
\section{Spectroscopy}
%
We adopt spectroscopic notation and denote (radially excited)
spin-parity eigenstates by $n^{2S+1}L_J$ (see table \ref{operators}).
\par
\begin{figure}
  \begin{center}
    \psfrag{m1}{\fbox{$h_b \left( ^1P_1\right)$}}
    \psfrag{l1}{$G_{1l}$}
    \psfrag{l2}{$G_{2l}$}
    \psfrag{Meff}{$\ln \frac{G(t+1)}{G(t)}$}
    \largepspicture{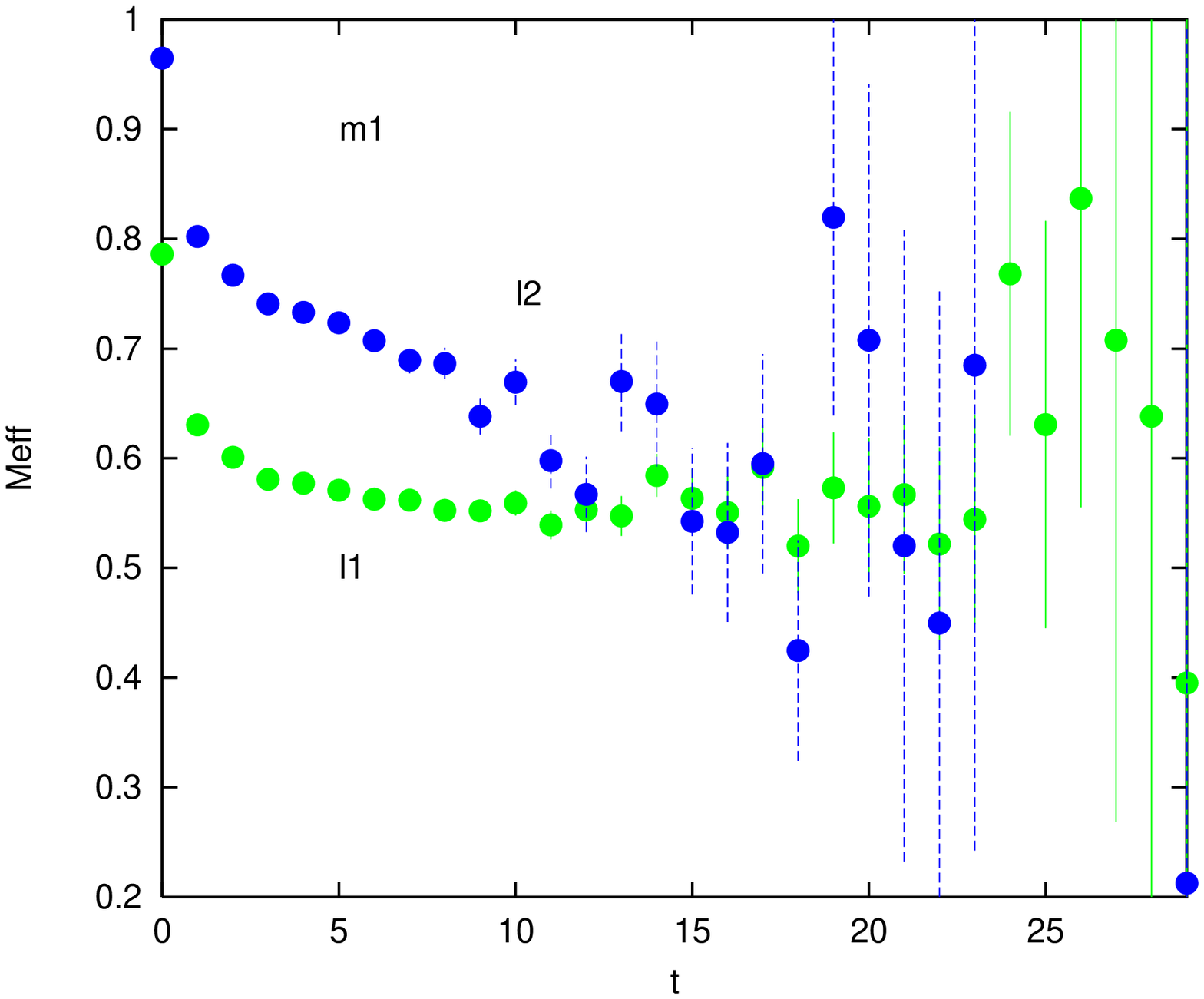}
  \end{center}
\caption{Two examples of effective masses with several radial
    excitations. The $\eta_b$ is at our lightest sea-quark, $\kappa =
    0.1575$ whereas the $h_b$ is at the heaviest, $\kappa = 0.156$.
\label{effmass}}
\end{figure}

%
%
Figure \ref{effmass} shows a subset of local masses: although we have
not tuned the smearing functions we generally find good plateaus for
ground and excited states. For $L=0$ the radially excited correlators
remain in the first/second  excitation for about ten timesteps. Local
masses  for P states are noisier and drop to the ground state without
staying in an `excited plateau' as nicely as for the $S$ states. 
%
\subsection{Fitting procedures}
%
To extract energies we fit
several smearing combinations of correlators simultaneously to a
multi-exponential ansatz. We find that vector fits to smeared-local
propagators, 
\begin{equation}
  G^{\rm meson}_{(sc,l)} (t) = \sum_{k=1}^{n_{exp}} b_{sc,k} {\rm
    e}^{-E_kt}\; ,
\label{fiteq}
\end{equation}
are quite stable whereas matrix fits appear to require higher statistics.
Energy levels are obtained from correlated
two-exponential fits to the two smearing functions $sc=1,2$ with eq. \ref{fiteq}.
Hyperfine splittings are determined from single exponential fits to the
ratio
\begin{equation}
  \frac{G^{\rm meson_A}_{(sc,sk)}}{G^{\rm meson_B}_{(sc,sk)}}
 (t) = A {\rm e}^{- (E_A - E_B) \cdot t}\; ,
\label{fiteqhyperfine}
\end{equation}
thus exploiting the strong
correlation between the data. We use smeared-local and smeared-smeared
correlators in eq.~\ref{fiteqhyperfine}. The results quoted correspond to the
value with the lowest $Q$-fit-parameter \cite{qdef} but with the statistical error
enlarged to take into account the uncertainty in the fitting ranges,
thus encompassing several possible fit-results from different fit-ranges.
For the hyperfine splittings the errors allow for the spread in
fit-results obtained using smeared-local as well as smeared-smeared
data. More complicated fit-functions confirm the
results obtained from the single exponential ansatz but do not behave
as stable. Results of our fits are given in tables \ref{tab_spec}
 and \ref{tab_splittings}.
%
%
\subsection{Light quark  mass dependence}
%
We now turn to the dependence of splittings on the dynamical quark
mass. Following the analysis in ref.~\cite{Grinstein} we extrapolate energy level
splittings linearly in the quark mass:
\begin{equation}
\Delta m = \Delta m_0  + c \sum_{u,d,s} m_q  \, ,
\end{equation}
where $m_q$ denotes the bare light quark masses. The relevant scale we pick is
given by $\frac{m_u + m_d + m_s}{3} \approx \frac{m_s}{3}$
corresponding to a ``physical pion mass'' of $\frac{m_\pi^2 + 2
  m_K^2}{3}$ \cite{Grinstein,Junko}. Since gluon momenta inside the
$b\bar b$ are much larger than any of the three lightest quark masses,
$\Upsilon$-physics should be insensitive to the $m_{u,d}-m_s$
mass-splitting. The value of the strange quark mass is taken from our
recent light quark mass calculation \cite{sesamlight} and is such that
\begin{equation}
m_{\rm eff} = \frac{m_s}{3} = \frac{1}{6a}\left(\frac{1}{\kappa_s} -
 \frac{1}{\kappa_c} \right) \simeq 0.0156
\end{equation}
 lies close to our lightest sea-quark mass, $a m_q(\kappa =
0.1575)$\footnote{The three sea-quark values correspond to pions
  $am_\pi = \{0.445(2), 0.341(3), 0.276(5)\}$ \cite{sesamspec}.}. 
The authors of \cite{Grinstein} have noted that
extrapolating to $m_{\rm eff} = \frac{m_s}{3}$ is appropriate for the
3-flavour case. In this context, we point out that there is an uncertainty
in the strange quark mass $m_s$ due to the fact that we have simulated
only two (degenerate) dynamical flavours. These are identified with the
up and down quarks so that, effectively, $m_s$ is defined in the
2-flavour theory in a ``partially quenched'' way only
\cite{sesamlight}. Thus, it is not clear what precise value of
$m_{\rm eff}$ mass splittings should be extrapolated to.
\par
\begin{figure}
  \begin{minipage}{6cm}
    \begin{center}
    \psfrag{mq}{$m_q$}
    \psfrag{m1}{{\small $1\bar P - 1^3S_1$}}
    \psfrag{m2}{{\small $2^3S_1 - 1^3S_1$}}
      \includegraphics[scale=0.6]{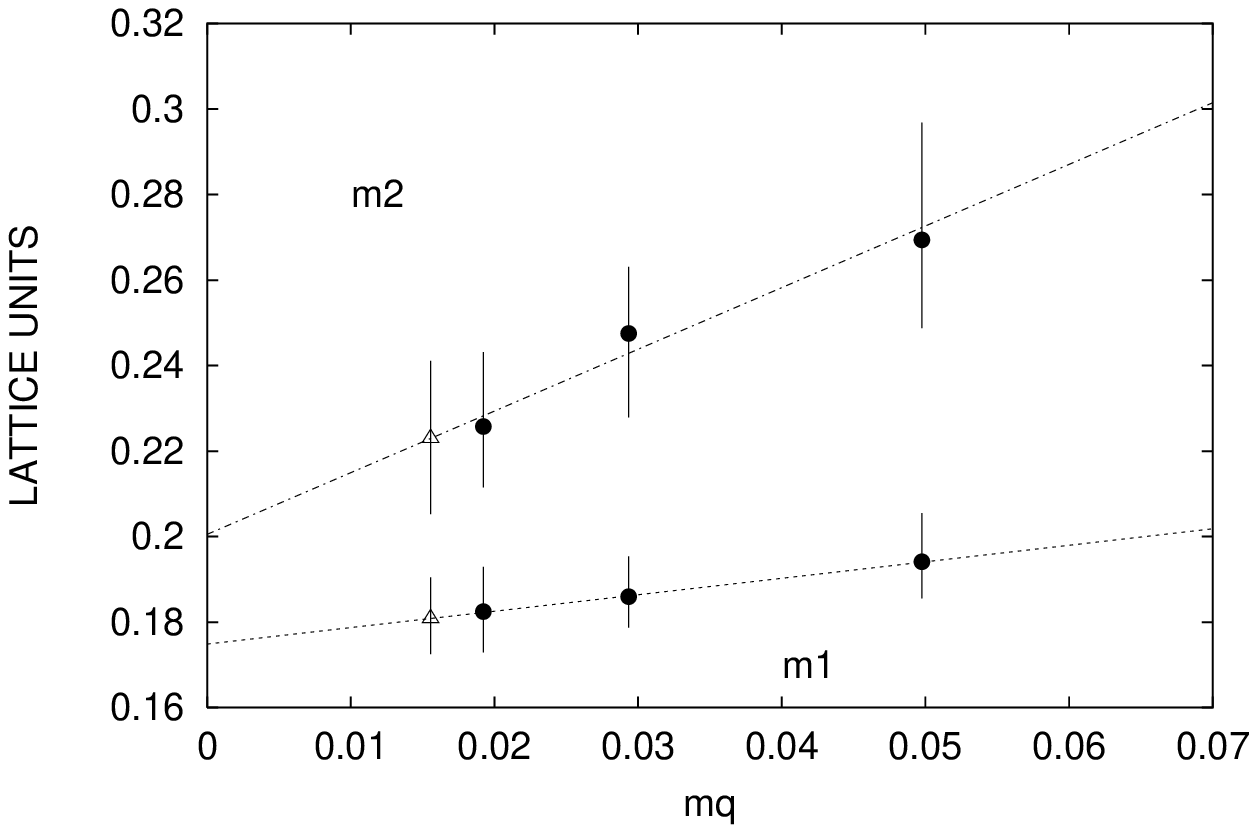}
    \end{center}
  \end{minipage}
  \hspace{2cm}
  \begin{minipage}{6cm}
    \begin{center}
    \psfrag{mq}{$m_q$}
    \psfrag{m1}{{\small $2^1P_1 - 1^1P_1$}}
    \psfrag{m2}{{\small $1\bar P - 1^3S_1$}}
      \includegraphics[scale=0.6]{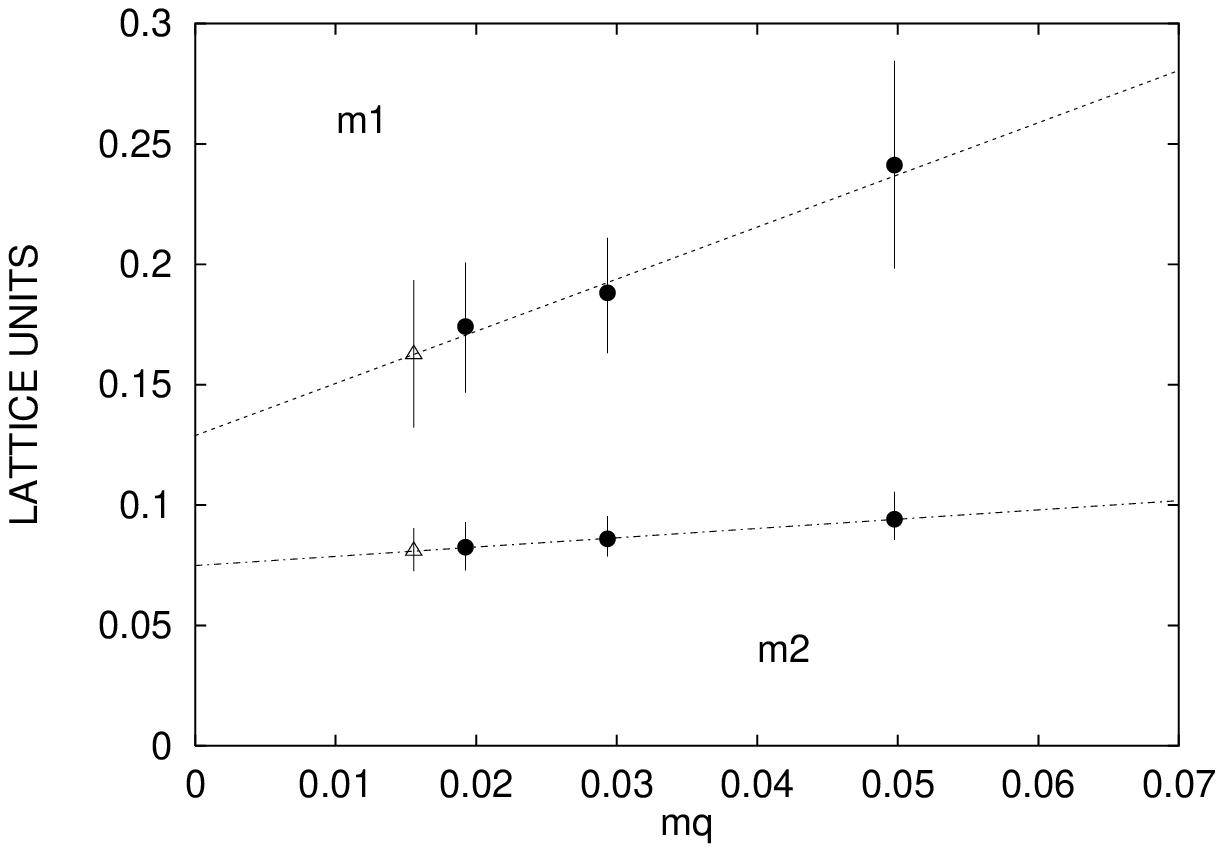}
    \end{center}
  \end{minipage}
  \caption{Extrapolation of spin independent splittings to
    $m_s/3$. The $1\bar P - 1^3S_1$ splitting is plotted for 
    comparison as it is one of the splittings used to determine 
    the scale (it is shifted downward by
    0.1 on the right plot). The triangular symbol denotes the
    extrapolated value.\label{chiral1}}
\end{figure}
\begin{figure}
  \begin{minipage}{6cm}
    \begin{center}
    \psfrag{mq}{$m_q$}
    \psfrag{m1}{{\small $1^3S_1 - 1^1S_0$}}
      \includegraphics[scale=0.6]{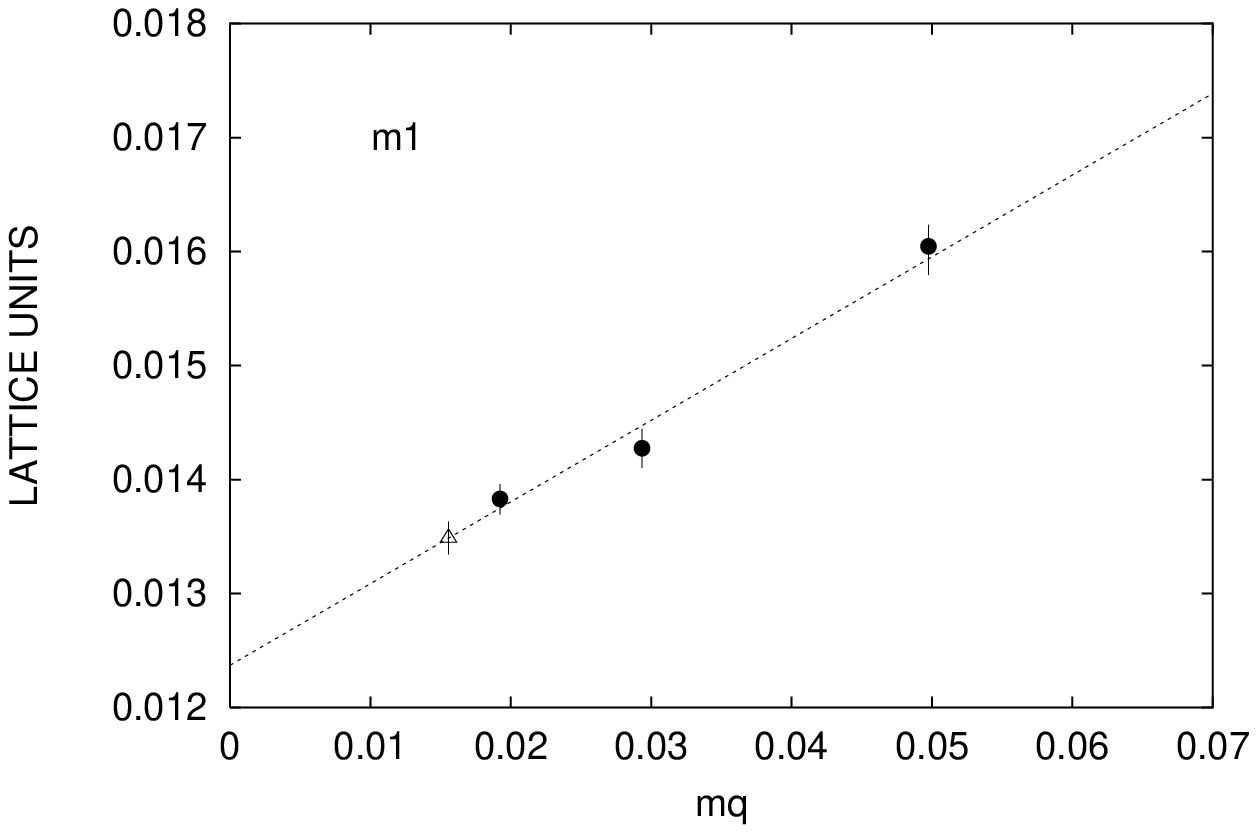}
    \end{center}
  \end{minipage}
  \hspace{2cm}
  \begin{minipage}{6cm}
    \begin{center}
    \psfrag{mq}{$m_q$}
    \psfrag{m1}{{\small $1^1P_1 - 1^3P_0$}}    
      \includegraphics[scale=0.6]{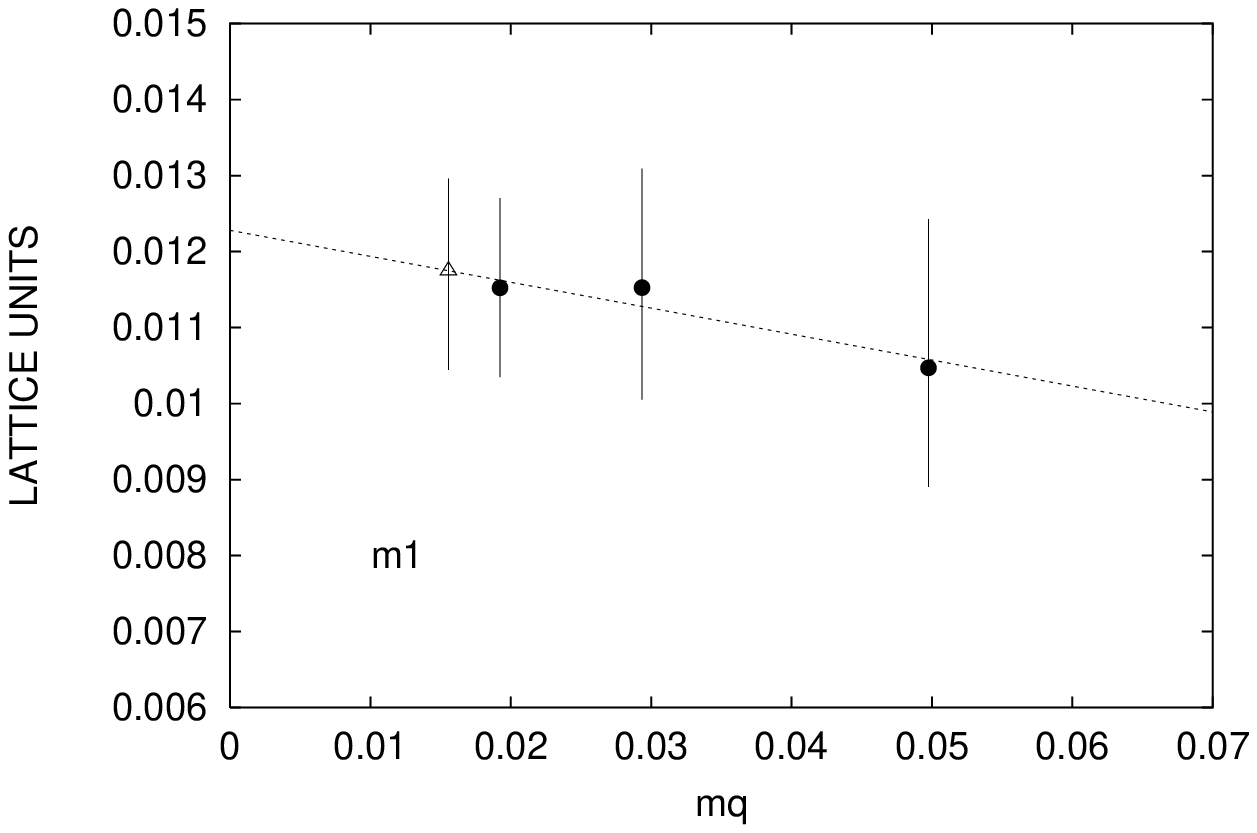}
    \end{center}
  \end{minipage}
\caption{Extrapolation of spin splittings to $m_s/3$. The triangular
    symbol denotes the extrapolated value.\label{chiral2}}
\end{figure}

%
%
%
%
The extrapolations are shown for various splittings in
figures \ref{chiral1} and \ref{chiral2}, table \ref{chiral_splittings}
contains $\Delta m(m_s/3)$ and $\Delta m_0$. No significant 
deviation from the linear parametrization is apparent in 
the data. With the exception of the $^3S_1 - {^1S_0}$
splitting the light quark dependence is too small to
be resolved given our current sensitivity. Within large errors,
we find $\frac{\Delta m(m_s/3)}{\Delta m_0}$, which can be taken as an
upper limit of the uncertainty in $m_{\rm eff}$, to be $3-20 \%$ for the spin-independent
splittings. Choosing $m_{\rm eff} < m_s/3$ would have the effect of
increasing the inverse lattice spacing $a^{-1}$ (see next
section). Clearly, much higher statistics is necessary to pin
down these effects.
\subsection{Lattice spacing}
The lattice scale is taken from the average of the $2^3S_1 - 1^3S_1$
and $CG(^3P)-1^3S_1$ splittings at $m_s/3$ (table
\ref{tab_spacings}). Recall that these splittings are virtually
insensitive to the heavy-quark mass and vary only by a few percent
from the $b \bar b$ to the $c \bar c$\footnote{The $2^3S_1 - 1^3S_1$ splittings
  are 562.9 and 589.1 for the $b \bar   b$ and $c \bar c$
  respectively; for the $CG(^3P)-1^3S_1$ they are 439.8   ($b \bar b$) and
  428.4 ($c \bar c$) - all in {\sf     MeV} \cite{barnett}.}. Note
that the average lattice spacing for $\beta = 5.6$ for $n_f=2$ agrees
very  well with the quenched one at $\beta = 6.0$ so that we can
compare our results in the ``full'' and in the quenched theories. 
\par
\begin{figure}
\begin{minipage}{6cm}
\begin{center}
 \psfrag{mq}{$m_q$}
      \includegraphics[scale=0.6]{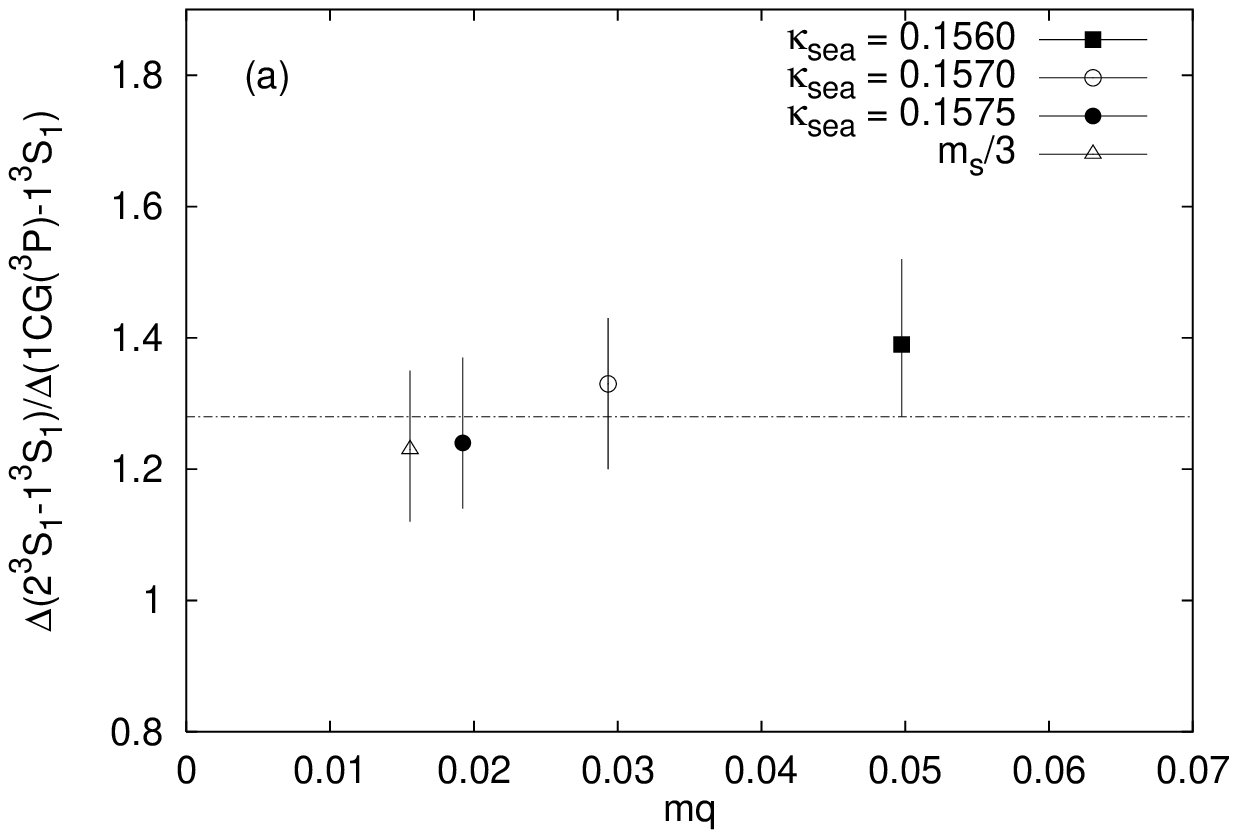}
\end{center}
\end{minipage}
\hspace{2cm}
\begin{minipage}{6cm}
\begin{center}
 \psfrag{nf}{$n_f$}
      \includegraphics[scale=0.6]{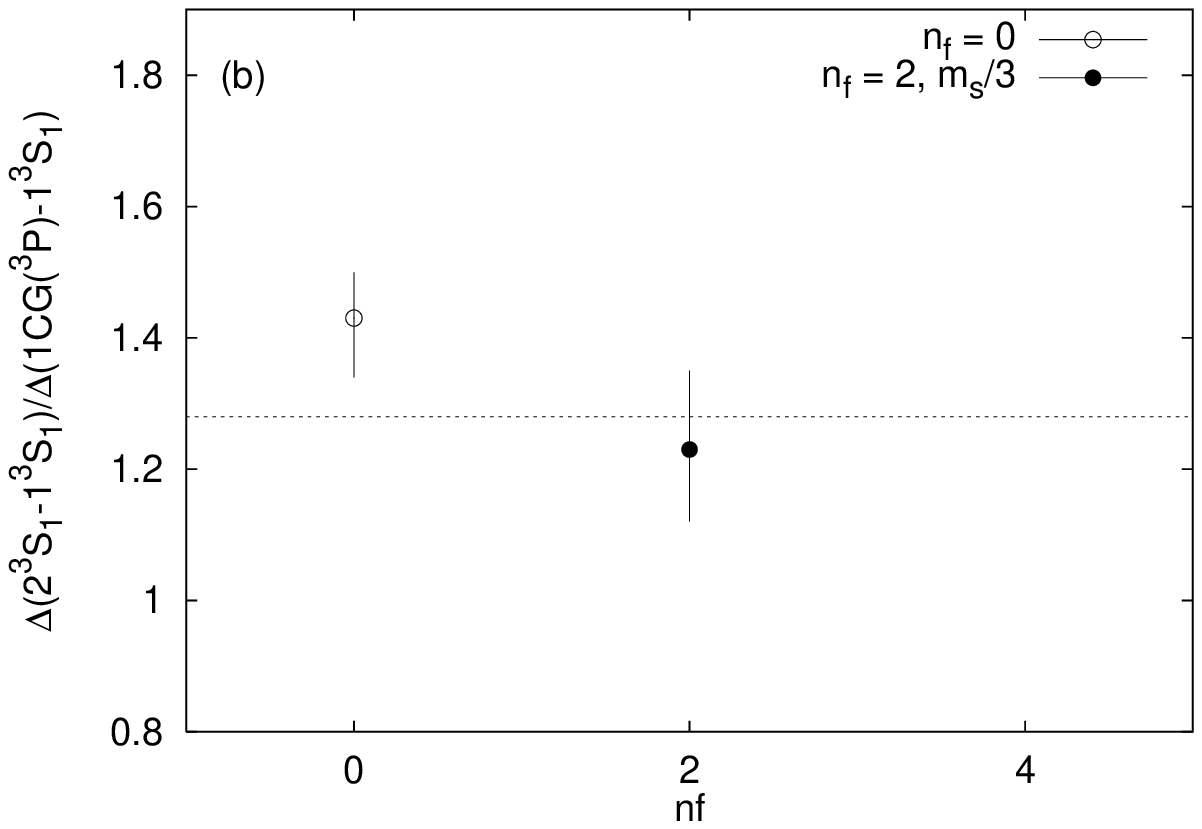}
\end{center}
\end{minipage}
\caption{The ratio of splittings $2^3S_1-1^3S_1$ to $1\bar P-1^3S_1$ as a
function of dynamical quark mass and number of dynamical quark
flavours.\label{fig:R_sp}}
\end{figure}

The ratio of splittings, $R_{\sf SP} = (2^3S_1 -
1^3S_1)/(CG(^3P) - 1^3S_1)$ is shown in figure \ref{fig:R_sp}(a) as a
function of the light quark mass. All values agree with the
experimental number of 1.28. Finally, we use the average lattice
spacings - at $m_s/3$ in the full theory - to convert the lattice
numbers to their physical values, see table \ref{tab_results}.  

%
%
%
%
%
%
%
\section{Discussion}
\begin{figure}[tb]
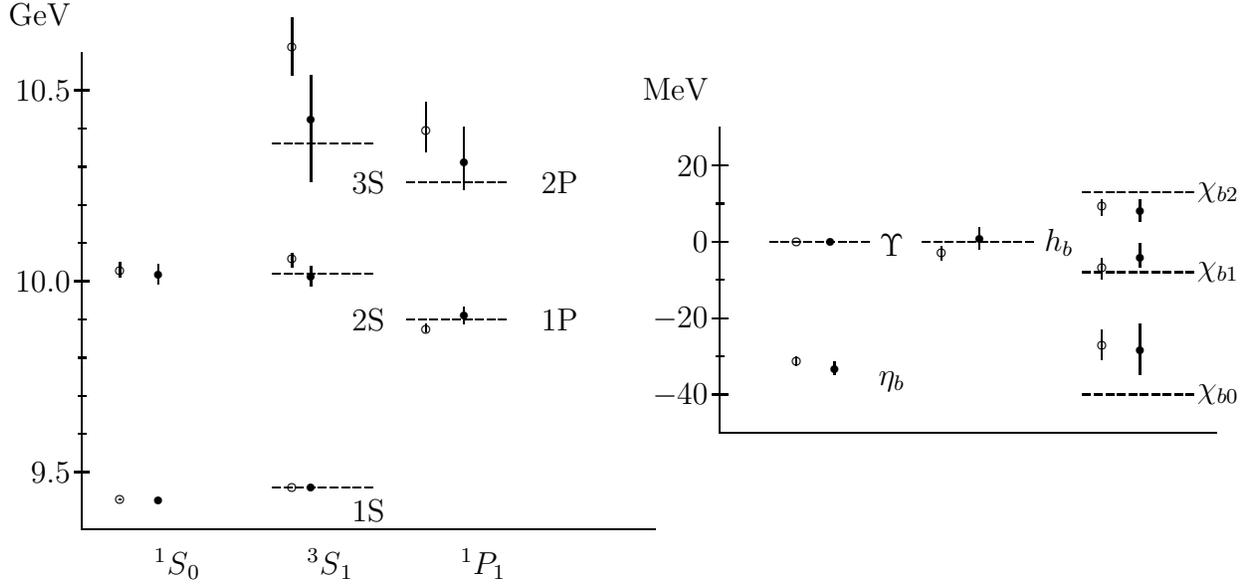

\begin{minipage}{6cm}
\begin{center}
\setlength{\unitlength}{.02in}
\begin{picture}(130,140)(10,930)
\put(15,935){\line(0,1){125}}
\multiput(13,950)(0,50){3}{\line(1,0){4}}
\multiput(14,950)(0,10){10}{\line(1,0){2}}
\put(12,950){\makebox(0,0)[r]{9.5}}
\put(12,1000){\makebox(0,0)[r]{10.0}}
\put(12,1050){\makebox(0,0)[r]{10.5}}
\put(12,1070){\makebox(0,0)[r]{GeV}}
\put(15,935){\line(1,0){150}}


\put(40,930){\makebox(0,0)[t]{${^1S}_0$}}



%

\input{plot_q60_11S0}
\input{plot_q60_21S0}


\input{plot_extrap_21S0}
\input{plot_extrap_11S0}


\put(80,930){\makebox(0,0)[t]{${^3S}_1$}}
\put(90,940){\makebox(0,0){1S}}
\multiput(65,946)(3,0){9}{\line(1,0){2}}
\put(90,990){\makebox(0,0){2S}}
\multiput(65,1002)(3,0){9}{\line(1,0){2}}
\put(90,1026){\makebox(0,0){3S}}
\multiput(65,1036)(3,0){9}{\line(1,0){2}}

\put(70,946){\circle{2}}
\input{plot_q60_23S1}
\input{plot_q60_33S1}

\put(75,946){\circle*{2}}
\input{plot_extrap_23S1}
\input{plot_extrap_33S1}


\put(120,930){\makebox(0,0)[t]{$^1P_1$}}

\put(140,990){\makebox(0,0){1P}}
\multiput(100,990)(3,0){9}{\line(1,0){2}}
\put(140,1026){\makebox(0,0){2P}}
\multiput(100,1026)(3,0){9}{\line(1,0){2}}

\input{plot_q60_11P1}
\input{plot_q60_21P1}

\input{plot_extrap_11P1}
\input{plot_extrap_21P1}

\end{picture}
\end{center}
\end{minipage}
\hspace{2cm}
\begin{minipage}{6cm}
%
%
%
%
\begin{center}
\setlength{\unitlength}{.02in}
\begin{picture}(100,80)(15,-50)
\put(15,-50){\line(0,1){80}}
\multiput(13,-40)(0,20){4}{\line(1,0){4}}
\multiput(14,-40)(0,10){7}{\line(1,0){2}}
\put(12,-40){\makebox(0,0)[r]{$-40$}}
\put(12,-20){\makebox(0,0)[r]{$-20$}}
\put(12,0){\makebox(0,0)[r]{$0$}}
\put(12,20){\makebox(0,0)[r]{$20$}}
\put(12,40){\makebox(0,0)[r]{MeV}}
\put(15,-50){\line(1,0){130}}

\multiput(28,0)(3,0){9}{\line(1,0){2}}
\put(60,2){\makebox(0,0)[t]{$\Upsilon$}}
\put(60,-34){\makebox(0,0)[t]{$\eta_b$}}


\put(35,0){\circle{2}}
\input{plot_q60_13S1-11S0}

\put(45,0){{\makebox(0,0){\circle*{2}}}}
\input{plot_extrap_13S1-11S0}

\multiput(68,0)(3,0){10}{\line(1,0){2}}
\put(100,0){\makebox(0,0)[l]{$h_b$}}

\input{plot_q60_1CG3P2-11P1}

\input{plot_extrap_1CG3P2-11P1}

\multiput(110,-40)(3,0){10}{\line(1,0){2}}
\put(140,-40){\makebox(0,0)[l]{$\chi_{b0}$}}
\multiput(110,-8)(3,0){10}{\line(1,0){2}}
\put(140,-8){\makebox(0,0)[l]{$\chi_{b1}$}}
\multiput(110,13)(3,0){10}{\line(1,0){2}}
\put(140,13){\makebox(0,0)[l]{$\chi_{b2}$}}

\input{plot_q60_1CG3P2-13P0}
\input{plot_q60_1CG3P2-13P1}
\input{plot_q60_13P2-1CG3P2}

\input{plot_extrap_1CG3P2-13P0}
\input{plot_extrap_1CG3P2-13P1}
\input{plot_extrap_13P2-1CG3P2}

\end{picture}
\end{center}
\end{minipage}
\vspace{0.8cm}
\caption{
The $b\bar b$ spectrum. The left plot shows radial splittings as well
as angular momentum splittings with the $\Upsilon$-level set to its
physical value. The right plot resolves the fine structure: here the
zero of energy is set to the $\Upsilon$-level in the left part and to
the spin averaged triplet P -level in the right part. Data points are
labeled as follows: open circles : $n_f = 0, \beta = 6.0$; filled
circles : $n_f = 2, m_q = m_s/3$.\label{fig:spectrum}}
\end{figure}

%
Figure \ref{fig:spectrum} summarizes our results for 2 and 0
flavours from table \ref{tab_results}. Unquenching effects are clearly
visible in the spin-independent part of the $b \bar b$ spectrum; this
is also evident from table \ref{tab_spacings} which shows that the
lattice spacings from the $1S-2S$ and the $1S-1P$ splittings do not
agree in the quenched theory whereas they coincide when 2
dynamical quarks are switched on. The ratio of these splittings is plotted in figure
\ref{fig:R_sp}(b) as a function of the number of flavours.
\par
We do not observe any significant impact of unquenching on the hyperfine
splittings. In particular, the P-hyperfine splittings seem to be
underestimated for both $n_f=0$ and $n_f=2$. Clearly, this result
needs to be corroborated with higher statistics. Errors on our
hyperfine splittings encompass the statistical error as well as the
uncertainty in the fitting range; the latter is essentially of the
same size as the statistical error so that we can expect some improvement
with higher statistics.
\par
\begin{figure}
\begin{minipage}{6cm}
\begin{center}
    \psfrag{m1}{{\small $1^3S_1 - 1^1S_0$}}    
      \includegraphics[scale=0.6]{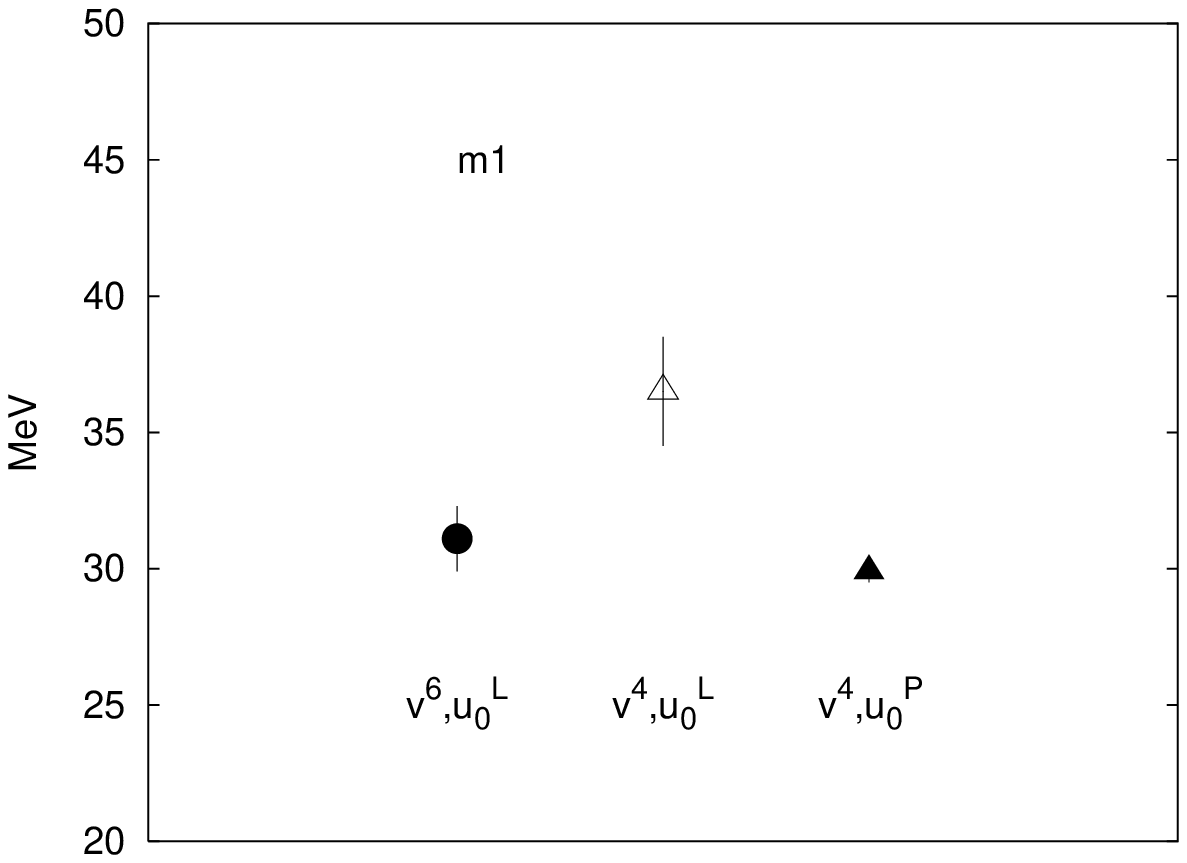}
\end{center}
\end{minipage}
\hspace{2cm}
\begin{minipage}{6cm}
\begin{center}
    \psfrag{m1}{{\small $2^3S_1 - 1^3S_1$}}    
      \includegraphics[scale=0.6]{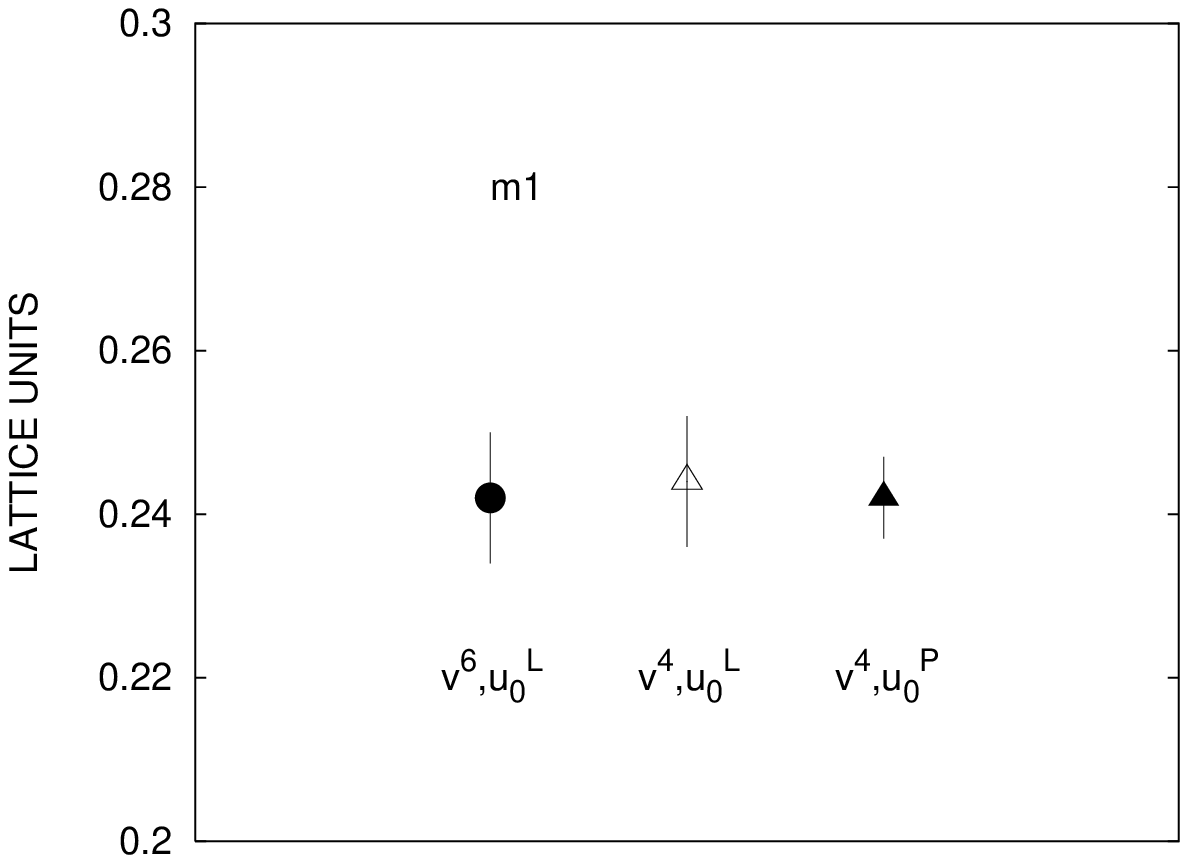}
\end{center}
\end{minipage}
\vspace{0.5cm}
\caption{On the left we compare the hyperfine S-splitting obtained
  with an ${\cal O}(M_b v^6)$ to the one obtained with an ${\cal
  O}(M_b v^4)$ correct NRQCD action. The higher order result is
  shifted significantly downward. Also shown (point on the very right)
  is the result obtained with an ${\cal O}(M_b v^4)$ action but using
  the plaquette tadpole scheme. The effect is to push the result down
relative to the one obtained with the Landau gauge mean link tadpole
scheme. On the right the same comparisons are shown for the
spin-independent $2S-1S$ splitting: as expected, no change is
seen. All results are for the quenched case with $\beta = 6.0$.
\label{figsystematics}}
\end{figure}

Compared to the only other simulation of NRQCD with dynamical quarks,
performed by the NRQCD collaboration\footnote{on gauge configurations
  generated by the HEMCGC group.} using staggered
fermions \cite{staggered,alpha}, 
we have introduced two
new features. Firstly, we have included all
spin-dependent corrections to ${\cal O}(M_b v^6)$ and secondly, we have
chosen an alternative way of removing the tadpole
diagrams\footnote{Other minor differences in the two simulations are:
  (i) a bare $b$-quark mass of 1.8 was used by the NRQCD collaboration; (ii) the HEMCGC
  data was not ``chirally'' extrapolated.}. Compared to 
an ${\cal O}(M_b v^4)$ correct action we expect the inclusion of ${\cal
  O}(M_b v^6)$ terms to effect the hyperfine splittings on the 10 \%
level, since $v^2 \approx 0.1$. Performing a relatively cheap ${\cal
  O}(M_b v^4)$ quenched simulation at $\beta = 6.0$ (with 200
measurements) and calculating only the $^3S_1 - {^1S_0}$ and the
(spin-independent) $2S-1S$ splittings we find this naive expectation
to be very well satisfied as is shown in figure \ref{figsystematics}.
%
%
%
Very recently, a similar effect was reported in \cite{manke} (${\cal
  O}$(10 \%) for the $b \bar b$) as well as in 
\cite{trottier}, where the effect of next-to-leading order
spin-dependent interactions for charmonium was found to be of ${\cal
  O}$(60 \%). Ref. \cite{manke} also found a rather severe shift, due
  to the ${\cal O}(M_b v^6)$ terms, of ${\cal O}$(30 \%) in the P-fine
  structure of the $b \bar b$. These results confirm Trottier's
  conclusion \cite{trottier} that the first three terms in the
  expansion for the hyperfine splitting are oscillating in
  sign. Therefore, the use of an ${\cal O}(M_b v^6)$-correct action
  is mandatory to get meaningful results for the fine-structure of
  the $\Upsilon$. 
\\
In addition to the order in the
NRQCD expansion, the hyperfine splittings will be sensitive to the
value of $u_0$. Naively, one expects the splitting to vary as
$1/u_0^4$ (since the $B$ fields contain four links) and the ratio of
the $u_0^4$ with the plaquette prescription to that with the mean link
Landau gauge value is roughly 10 \% (see table \ref{tadpole} for values of
$u_0$ in both schemes). We therefore also plot
in figure \ref{figsystematics} the $^3S_1 - {^1S_0}$ splitting obtained with  the
${\cal O}(M_b v^4)$ correct action but using the plaquette prescription
for $u_0$ (this result is from the NRQCD collaboration\footnote{We have reproduced
  their value on 200 configurations but with a larger error.}). The
plot confirms the naive expectation and shows that changing $u_0$ to
the mean link Landau value shifts the hyperfine splitting in the {\it
  opposite} direction as adding the 
${\cal O}(M_b v^6)$ spin-dependent corrections. For reference the
corresponding value is also shown in the spin-independent $2S-1S$
splitting where, within the larger error, no effect is seen. 
\par
\begin{figure}[tb]
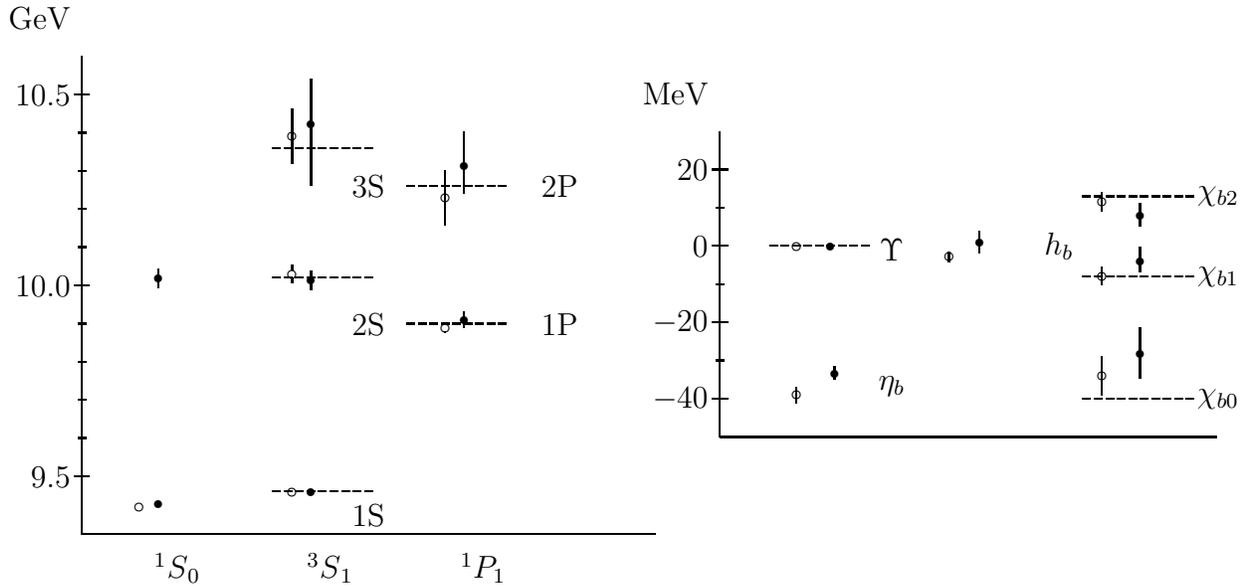

\begin{minipage}{6cm}
\begin{center}
\setlength{\unitlength}{.02in}
\begin{picture}(130,140)(10,930)
\put(15,935){\line(0,1){125}}
\multiput(13,950)(0,50){3}{\line(1,0){4}}
\multiput(14,950)(0,10){10}{\line(1,0){2}}
\put(12,950){\makebox(0,0)[r]{9.5}}
\put(12,1000){\makebox(0,0)[r]{10.0}}
\put(12,1050){\makebox(0,0)[r]{10.5}}
\put(12,1070){\makebox(0,0)[r]{GeV}}
\put(15,935){\line(1,0){150}}


\put(40,930){\makebox(0,0)[t]{${^1S}_0$}}


\put(30,942){\circle{2}}



\input{plot_extrap_21S0}
\input{plot_extrap_11S0}


\put(80,930){\makebox(0,0)[t]{${^3S}_1$}}
\put(90,940){\makebox(0,0){1S}}
\multiput(65,946)(3,0){9}{\line(1,0){2}}
\put(90,990){\makebox(0,0){2S}}
\multiput(65,1002)(3,0){9}{\line(1,0){2}}
\put(90,1026){\makebox(0,0){3S}}
\multiput(65,1036)(3,0){9}{\line(1,0){2}}

\put(70,946){\circle{2}}
\put(70,1003){\circle{2}}
\put(70,1004){\line(0,1){1.4}}
\put(70,1002){\line(0,-1){1.4}}
\put(70,1039.1){\circle{2}}
\put(70,1039.1){\line(0,1){7.2}}
\put(70,1039.1){\line(0,-1){7.2}}

\put(75,946){\circle*{2}}
\input{plot_extrap_23S1}
\input{plot_extrap_33S1}


\put(120,930){\makebox(0,0)[t]{$^1P_1$}}

\put(140,990){\makebox(0,0){1P}}
\multiput(100,990)(3,0){9}{\line(1,0){2}}
\put(140,1026){\makebox(0,0){2P}}
\multiput(100,1026)(3,0){9}{\line(1,0){2}}

\put(110,989){\circle{2}}
\put(110,990){\line(0,1){0.2}}
\put(110,988){\line(0,-1){0.2}}
\put(110,1023){\circle{2}}
\put(110,1023){\line(0,1){7.2}}
\put(110,1023){\line(0,-1){7.2}}

\input{plot_extrap_11P1}
\input{plot_extrap_21P1}

\end{picture}
\end{center}
\end{minipage}
\hspace{2cm}
%
%
%
%
\begin{minipage}{6cm}
\begin{center}
\setlength{\unitlength}{.02in}
\begin{picture}(100,80)(15,-50)
\put(15,-50){\line(0,1){80}}
\multiput(13,-40)(0,20){4}{\line(1,0){4}}
\multiput(14,-40)(0,10){7}{\line(1,0){2}}
\put(12,-40){\makebox(0,0)[r]{$-40$}}
\put(12,-20){\makebox(0,0)[r]{$-20$}}
\put(12,0){\makebox(0,0)[r]{$0$}}
\put(12,20){\makebox(0,0)[r]{$20$}}
\put(12,40){\makebox(0,0)[r]{MeV}}
\put(15,-50){\line(1,0){130}}

\multiput(28,0)(3,0){9}{\line(1,0){2}}
\put(60,2){\makebox(0,0)[t]{$\Upsilon$}}
\put(60,-34){\makebox(0,0)[t]{$\eta_b$}}

\put(35,0){\circle{2}}
\put(35,-39){\circle{2}}
\put(35,-39){\line(0,1){2}}
\put(35,-39){\line(0,-1){2}}


\put(45,0){{\makebox(0,0){\circle*{2}}}}
\input{plot_extrap_13S1-11S0}

\put(100,0){\makebox(0,0)[l]{$h_b$}}

\put(75,-2.9){\circle{2}}
\put(75,-2.9){\line(0,1){1.2}}
\put(75,-2.9){\line(0,-1){1.2}}

\input{plot_extrap_1CG3P2-11P1}

\multiput(110,-40)(3,0){10}{\line(1,0){2}}
\put(140,-40){\makebox(0,0)[l]{$\chi_{b0}$}}
\multiput(110,-8)(3,0){10}{\line(1,0){2}}
\put(140,-8){\makebox(0,0)[l]{$\chi_{b1}$}}
\multiput(110,13)(3,0){10}{\line(1,0){2}}
\put(140,13){\makebox(0,0)[l]{$\chi_{b2}$}}

\put(115,-34){\circle{2}}
\put(115,-34){\line(0,1){5}}
\put(115,-34){\line(0,-1){5}}
\put(115,-7.9){\circle{2}}
\put(115,-7.9){\line(0,1){2.4}}
\put(115,-7.9){\line(0,-1){2.4}}
\put(115,11.5){\circle{2}}
\put(115,11.5){\line(0,1){2.4}}
\put(115,11.5){\line(0,-1){2.4}}

\input{plot_extrap_1CG3P2-13P0}
\input{plot_extrap_1CG3P2-13P1}
\input{plot_extrap_13P2-1CG3P2}

\end{picture}
\end{center}
\end{minipage}
\vspace{0.5cm}
\caption{Comparison of our $\{{\cal O}(mv^6), u_0^{(L)}\}$
  Wilson-fermion result extrapolated to $m_s/3$ (full circles) to the $\{{\cal
  O}(mv^4), u_0^{(P)}\}$ (open circles) staggered fermion result at
  $m_q = 0.01$ of the NRQCD collaboration.\label{Wil_Kog}}
\end{figure}

With this in mind, we compare in figure \ref{Wil_Kog} our $\{{\cal
  O}(M_b v^6), u_0^{(L)}, n_f=2 \}$ results to the $\{{\cal
  O}(M_b v^4), u_0^{(P)}, n_f=2 \}$ of the NRQCD
  collaboration \cite{staggered}. Both in the spin-independent and in
  the spin-dependent sectors the results are in good agreement although
  our splittings tend to be slightly smaller.
\section{Summary and Outlook}
We have presented the first calculation of the $b \bar b$ spectrum
with dynamical Wilson fermions\footnote{We note that the authors of \cite{manke1} are
  presently starting a similar analysis.}. Our non-relativistic lattice action is
correct to ${\cal O}(M_b v^4)$ for spin-independent operators and
includes all spin-dependent corrections of ${\cal O}(M_b v^6)$. We rely
on tadpole improvement choosing the mean link in Landau gauge as our
improvement scheme.
\par
By use of appropriate smearing techniques we obtain clean signals for
ground and excited states.
\par
We have studied the light quark mass and flavour dependence of the $b
\bar b$ spectrum. All quantities
were linearly extrapolated to $m_s/3$. We are helped in these
extrapolations by the fact that $m_s/3$ is very close to
our lightest sea-quark mass so that the extrapolated values are easily
consistent within errors with the values at $\kappa = 0.1575$. The
lattice spacings were determined after ``chiral extrapolation'' 
with an error of 6 \%, which accounts for the statistical and systematic
(fitting) uncertainties. 
\par
Comparing to the quenched
calculation at similar lattice spacing we find a distinct unquenching
effect in the spin-independent splittings which can be quantified by
the ratio $R_{\rm SP} = \frac{2^3S_1 - 1^3S_1}{1 \bar P - 1^3S_1}$
determined to be $1.23(11)$ with 2 dynamical flavours and $1.43(9)$
for the quenched case; the experimental value is 1.28.
\par
We have estimated the systematic error arising due to two different
choices of the tadpole subtraction factor $u_0$; whereas
spin-independent quantities remain unaffected by changes in $u_0$
the fine splittings increase when using the mean link Landau gauge
$u_0$. Interestingly, we find this effect to be of the same
magnitude but of {\it opposite} direction as switching on the
spin-dependent ${\cal O}(M_b v^6)$ corrections. These two effects thus
combine to yield good agreement between our $\{{\cal  O}(M_b v^6),
u_0^{(L)}, n_f=2 \}$ results and  the $\{{\cal   O}(M_b v^4),
u_0^{(P)}, n_f=2\}$ results of the NRQCD collaboration with dynamical
staggered fermions \cite{staggered}.
\par
Our results confirm that the ${\cal O}(M_b v^6)$ terms in the action
are mandatory for calculating the hyperfine splittings. As expected,
the determination of the lattice spacing $a^{-1}$ (as used, for
example, in the determination of $\alpha_s$ \cite{alpha}) remains
unaffected by changing tadpole schemes. 
\par
Overall, NRQCD has proven to work very well in the spin-independent
sector, in particular in giving a precise determination of the lattice
spacing $a$. Given the weak dependence of these splittings on the
heavy quark (experimentally) as well as on the light quark masses, as
found here, it appears worthwhile to push the scale determinations to higher
statistics using our configurations generated by \tkl at $\kappa =
0.1575$ on $24^3 \times 40$ lattices ($\beta = 5.6$). This would
enable a reliable extrapolation to 3 or 4 flavours. The dominant
source of error in such a lattice scale determination is most likely
due to the remaining lattice discretisation errors in the gauge
configurations. 
\par
Although higher statistics are highly desirable, it seems unlikely that
progress in the spin-dependent sector will come from this
approach alone. It is worrying to find the spin-dependent corrections of
${\cal O}(M_b v^6)$ as large as 10 \% even for the $b
\bar b$, decreasing the fine-splittings relative to the lower
order estimates. Even at this order the NRQCD expansion does not
describe the hyperfine structure of the spectrum. A perturbative or
non-perturbative calculation of the coefficients $c_i$ is badly
needed\footnote{Work in this direction was recently started by
  \cite{trottier_lat97}. Perturbative calculations \cite{Morningstar}
  indicate that radiative corrections to $c_1$ and $c_5$ are small.}.
\section*{Acknowledgements}
H.~H. and A.~S. wish to thank C.~Davies for a number of fruitful discussions. We
thank A.~Wachter for the kind donation of his Schr\"odinger equation
solver! 
\par
We appreciate support by EU network grant No.~ERB CHRX CT 92-0051. 
A.~S. acknowledges support of the DFG-Graduiertenkolleg. 
Part of the computations have been performed on the Connection Machine
CM-5 of the Institut for Angewandte Informatik, Wuppertal.

%
%

\begin{figure}
  \begin{center}
    \psfrag{m1}{\fbox{$\eta_b \left( ^1S_0\right)$}}
    \psfrag{l1}{$G_{1l}$}
    \psfrag{l2}{$G_{2l}$}
    \psfrag{l3}{$G_{3l}$}
    \psfrag{Meff}{$\ln \frac{G(t+1)}{G(t)}$}
    \largepspicture{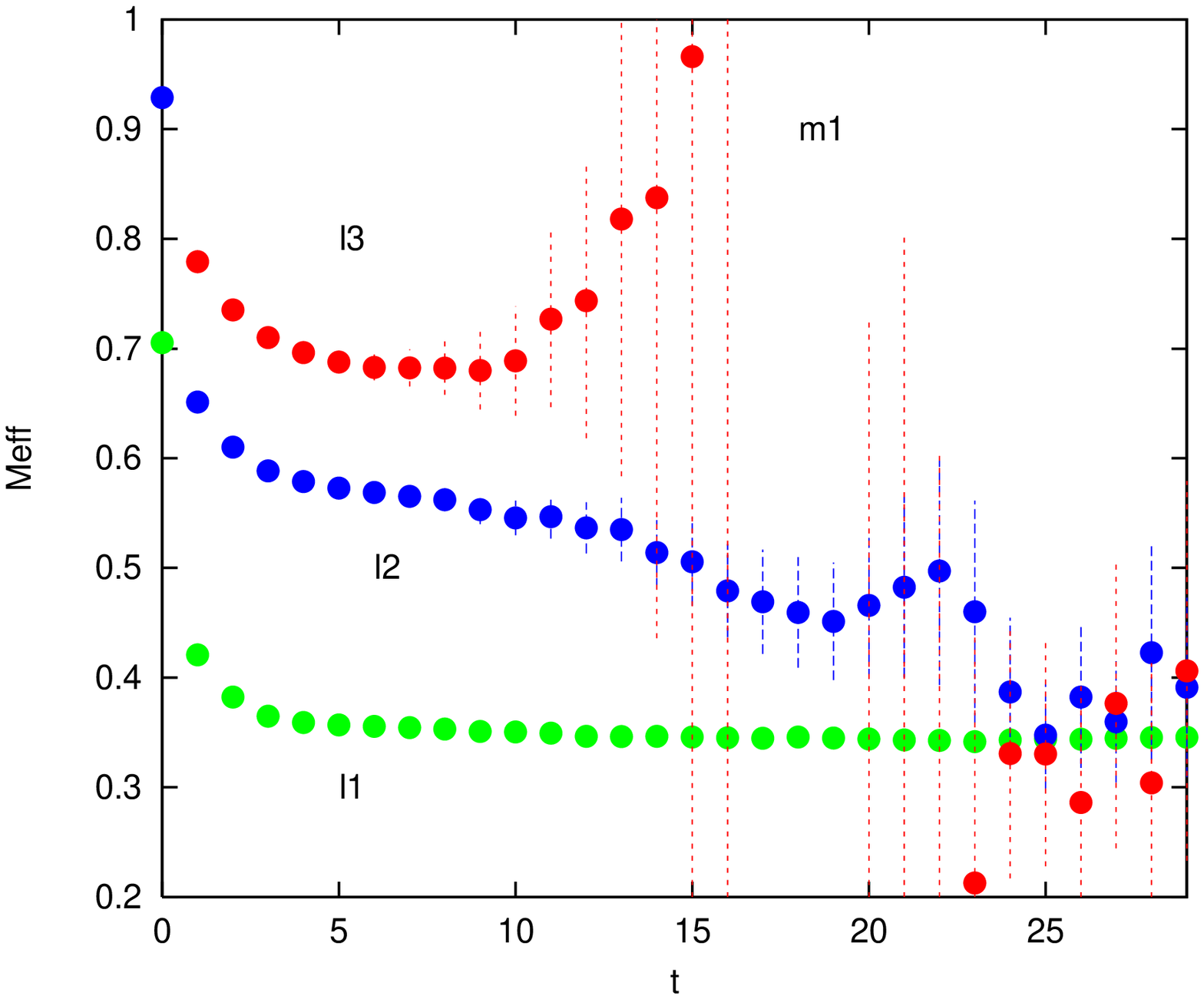}
  \end{center}
\end{figure}
\newpage
\vspace{2cm}
\newpage
%
%
%
%

%
%
%
%

%
%
%

\begin{table}
\caption{Comparison of tadpole improvement schemes.
\label{tadpole}}
\begin{center}
\begin{tabular}{ccccc}
$u_0$ & $\kappa = 0.156$ &$\kappa = 0.157$& $\kappa = 0.1575$ & quenched \\
\hline \hline
$u_0^{(P)}=\sqrt[4]{\langle\frac{1}{3}Tr U_{\mu \nu}\rangle}$ & $0.8688$ &
$0.8695$ & $0.8697$ & $0.8778$ \\
$u_0^{(L)}=\langle\frac{1}{3}Tr U_{\mu}\rangle$ & $0.8499$ & $0.8519$ & $0.8529$ & 
$0.8608$\\
\end{tabular}
\end{center}
\end{table}

\begin{table}
\caption{Operators and naming scheme.
\label{operators}}
\begin{center}
\begin{tabular}{ccc}
$b \bar b$ & $^{2S + 1}L_J (J^{PC})$ & $\Omega$ \\
\hline \hline
$\eta_b$ & $^1S_0 (0^{-+})$ & $1$ \\
\hline
$\Upsilon$ & $^3S_1(1^{--})$ & $\sigma_i$ \\
\hline
$h_b$ & $^1P_1(1^{--})$ & $\Delta_i$ \\
\hline
$ \chi_{b_0}$ & $^3P_0(0^{++})$ & $\sum_{j=1}^3 \Delta_j \sigma_j$ \\
\hline
$ \chi_{b_1}$ & $^3P_1(1^{++})$ & $\Delta_i\sigma_j - \Delta_j \sigma_i$ \\
\hline
$ \chi_{b_2}$ & $^3P_2(2^{++})$ & $\Delta_i \sigma_i - \Delta_j
\sigma_j$ \\
              &                 & $\Delta_i \sigma_j + \Delta_j
              \sigma_i$ \\   
\hline
$\bar \chi $ & $CG(^3P) \equiv \overline P = \frac{5^3P_2 + 3 ^3P_1 + ^3P_0}{9}$ & - \\
\end{tabular}
\end{center}
\end{table}

\begin{table}
\caption{Simulation details ($n_f$ denotes the number of flavours).
\label{simulation}}
\begin{center}
\begin{tabular}{cccc}
\multicolumn{4}{c} {$\beta_{\rm dyn} = 5.6$, $n_f = 2$, $16^3 \times
  32$}\\
\hline
$\kappa_{\rm sea}$ & 0.156 & 0.1570 & 0.1575 \\ 
\hline
number of configurations & 200 & 200 & 200 \\   
\hline
measurements & 800 & 800 & 800 \\
\hline
\hline
\multicolumn{4}{c} {$\beta_{\rm quenched} = 6.0$, $n_f = 0$, $16^3 \times 32$}\\
\hline
\multicolumn{4}{c}{number of configurations: 200}\\
\hline
\multicolumn{4}{c}{measurements: 800}\\
\end{tabular}
\end{center}
\end{table}

\begin{table}
\caption{Spectrum results in lattice units for all three sea-quark masses and for the
  quenched simulation.
\label{tab_spec}}
\begin{center}
\begin{tabular}{ccccc}
& \multicolumn{3}{c}{$n_f = 2$, $\beta = 5.6 $} & $n_f = 0$, $\beta = 6.0
$ \\
\hline
Level & $\kappa = 0.1560 $ & $\kappa = 0.1570 $ & $\kappa = 0.1575
$ & \\
\hline
\hline
$ 1^1S_0 $ & 0.3493(8) & 0.3476(7) & 0.3446(7) &  0.3309(8) \\
$ 2^1S_0 $ & 0.635(20) & 0.613(15)  & 0.588(18)&  0.574(12)  \\
$ 1^3S_1 $ & 0.3651(9) & 0.3621(9) & 0.3582(9) &  0.3438(8) \\
$ 2^3S_1 $ & 0.635(21) & 0.610(15) & 0.584(17) &  0.586(13) \\
$ 3^3S_1 $ &           &           & 0.75(5)   &  0.81(2)   \\
$ 1^1P_1 $ & 0.559(11) & 0.548(9)  & 0.541(10) &  0.512(9)  \\
$ 2^1P_1 $ & 0.80(4)   & 0.74(2)   & 0.71(2)   &  0.72(2)   \\
\end{tabular}
\end{center}
\end{table}

\begin{table}
\caption{Splittings at fixed sea-quark as well as quenched
  results (in lattice units).
\label{tab_splittings}}
\begin{center}
\begin{tabular}{ccccc}
& \multicolumn{3}{c}{$n_f = 2$, $\beta = 5.6 $} & $n_f = 0$, $\beta = 6.0
$ \\
\hline
Splitting & $\kappa = 0.1560 $ & $\kappa = 0.1570 $ & $\kappa = 0.1575$ &  \\
\hline
\hline
$ 2^1S_0 - 1^1S_0 $ & 0.285(20) & 0.266(9) & 0.243(7)   & 0.243(8) \\
$ 2^3S_1 - 1^3S_1 $ & 0.269(25) & 0.248(15)& 0.226(17)  & 0.242(13)\\ 
$ 3^3S_1 - 1^3S_1 $ &           &          & 0.39(6)    & 0.47(2)  \\
$ 1^1P_1 - 1^3S_1 $ & 0.194(10) & 0.186(9) & 0.183(10)  & 0.168(9) \\
$ 2^1P_1 - 1^1P_1 $ & 0.24(4)   & 0.19(2)  & 0.17(2)    & 0.21(2)  \\
$ 1\bar P - 1^3S_1 $ & 0.194(11)& 0.186(9) & 0.182(10)  & 0.169(9) \\ 
\hline
\hline
$ 1^3S_1 - 1^1S_0 $ & 0.0160(2)  & 0.0143(2) & 0.0138(2)  & 0.0126(2) \\
$ 1^3P_2 - 1^1P_1 $ & 0.0046(10) & 0.0031(10)& 0.0032(10) & 0.0050(6) \\
$ 1^1P_1 - 1^3P_1 $ & 0.0023(11) & 0.0028(9) & 0.0017(6)  & 0.0016(8) \\
$ 1^1P_1 - 1^3P_0 $ & 0.0105(19) & 0.0115(15)& 0.0115(11) & 0.0098(8) \\
\hline
\hline
$ 1^3P_2 - 1\bar P $ & 0.0039(8) & 0.0036(7) & 0.0033(5)  & 0.0038(5) \\
$ 1\bar P - 1^3P_1 $ & 0.0029(9) & 0.0023(7) & 0.0017(6)  & 0.0027(7) \\
$ 1\bar P - 1^3P_0 $ & 0.0111(15)& 0.0110(13)& 0.0115(11) & 0.0110(9) \\
$ 1\bar P - 1^1P_1 $ & 0.0006(7) & -0.0005(7)& -0.0001(5) & 0.0011(4) \\
\end{tabular}
\end{center}
\end{table}

\begin{table}
\caption{Splitting results (in lattice units) extrapolated to
  $m_s/3$.
\label{chiral_splittings}}
\begin{center}
\begin{tabular}{ccc}
Splitting & $ \Delta m(m_s/3) $ &  $\Delta m_0$ \\
\hline
\hline
$ 2^1S_0 - 1^1S_0 $ & 0.238(9)  & 0.212(18) \\
$ 2^3S_1 - 1^3S_1 $ & 0.223(18) & 0.201(29)\\ 
$ 1^1P_1 - 1^3S_1 $ & 0.181(9)  & 0.176(16) \\
$ 2^1P_1 - 1^1P_1 $ & 0.16(3)   & 0.13(5)  \\
$ 1\bar P - 1^3S_1$ & 0.181(9)  & 0.175(16) \\ 
\hline              
\hline              
$ 1^3S_1 - 1^1S_0 $ & 0.0135(2)  & 0.0124(3) \\
$ 1^3P_2 - 1^1P_1 $ & 0.0029(9) & 0.0021(15) \\
$ 1^1P_1 - 1^3P_1 $ & 0.0019(7) & 0.0015(13) \\
$ 1^1P_1 - 1^3P_0 $ & 0.0117(13)& 0.0123(22) \\
\hline              
\hline              
$ 1^3P_2 - 1\bar P$ & 0.0032(6) & 0.0029(9) \\
$ 1\bar P - 1^3P_1$ & 0.0016(7) & 0.0010(11) \\
$ 1\bar P - 1^3P_0$ & 0.0114(12)& 0.0116(21) \\
$ 1\bar P - 1^1P_1$ & -0.0003(5)& -0.0007(10) \\
\end{tabular}
\end{center}
\end{table}

\begin{table}
\caption{Determination of the lattice spacing from the $2^3S_1-1^3S_1$ and
  $CG(^3P)-1^3S_1$ splittings. We use the average value to convert our
  results to physical units. $R_{\sf SP}$ is to be compared to the
  experimental value of 1.28.\label{tab_spacings}}
\begin{center}
\begin{tabular}{ccccc}
& Splitting & $a^{-1}[{\sf GeV}]$ & Average $a^{-1}$[{\sf GeV}] &
$R_{\sf SP}$ \\
\hline\hline
$n_f = 0$, $\beta = 6.0$ & $\Upsilon ' - \Upsilon$ & 2.33(10) &
2.47(10) & 1.43(9) \\
& $\bar\chi - \Upsilon $ & 2.61(13) & & \\
\hline
$\kappa = 0.1560 $ & $\Upsilon ' - \Upsilon$ & 2.09(18) & 2.18(12) &
1.39(13) \\
& $\bar\chi - \Upsilon $ & 2.27(12) & & \\
\hline 
$\kappa = 0.1570 $ & $\Upsilon ' - \Upsilon$ & 2.27(15) & 2.32(10) &
1.33(12) \\
& $\bar\chi - \Upsilon $ & 2.36(10) & & \\
\hline 
$\kappa = 0.1575 $ & $\Upsilon ' - \Upsilon$ & 2.49(17) & 2.45(12) &
1.24(13) \\
& $\bar\chi - \Upsilon $ & 2.41(10) & & \\
\hline
$ m_s/3 $ & $\Upsilon ' - \Upsilon$ & 2.52(20) & 2.48(14) &
1.23(11) \\
& $\bar\chi - \Upsilon $ & 2.43(12) & & \\
\end{tabular}
\end{center}
\end{table}

\begin{table}
\caption{Overview of our results in physical units for full QCD and
  the quenched case.\label{tab_results}}
\begin{center}
\begin{tabular}{cccc}
& \multicolumn{2}{c}{Simulation Result [{\sf GeV}]} & Experiment\cite{barnett} \\
\hline
Splitting & $ m_s/3 $ & $n_f=0$ , $\beta = 6.0 $ & \\
\hline
\hline
$ 2^1S_0 - 1^1S_0 $ & 0.591(26) & 0.600(20) & \\
$ 2^3S_1 - 1^3S_1 $ & 0.553(25) & 0.598(20) & 0.5629 \\
$ 3^3S_1 - 1^3S_1 $ &           & 1.15(7)   & 0.895 \\
$ 1^1P_1 - 1^3S_1 $ & 0.450(22) & 0.416(13) & \\
$ 2^1P_1 - 1^1P_1 $ & 0.40(8)   & 0.52(7)   &  \\
$ 1\bar P - 1^3S_1 $ & 0.448(22)& 0.419(12)& 0.4398 \\ 
\hline
\hline
$ 1^3S_1 - 1^1S_0 $ & 0.0334(19) & 0.0313(12) & \\
$ 1^3P_2 - 1^1P_1 $ & 0.0071(25) & 0.0122(15) & \\
$ 1^1P_1 - 1^3P_1 $ & 0.0046(18) & 0.0039(20) & \\
$ 1^1P_1 - 1^3P_0 $ & 0.0291(30) & 0.0243(21) & \\
\hline
\hline
$ 1^3P_2 - 1\bar P $ & 0.0080(15) & 0.0094(12) & 0.0130 \\
$ 1\bar P - 1^3P_1 $ & 0.0040(18) & 0.0067(15) & 0.0083 \\
$ 1\bar P - 1^3P_0 $ & 0.0283(35) & 0.0271(20) & 0.040 \\
$ 1\bar P - 1^1P_1 $ & -0.0009(15)
& 0.0028(10) &        \\
\end{tabular}
\end{center}
\end{table}

\end{document}